\documentclass[useAMS,usenatbib]{mn2e}

\usepackage{graphicx}

\title[Galactic PNe magnetic alignment]
{On the alignment of PNe and local magnetic field at the galactic centre: MHD numerical simulations}
\author[D. Falceta-Gon\c{c}alves \& H. Monteiro]{D. Falceta-Gon\c{c}alves$^{1,2}$\thanks{E-mail:dfalceta@usp.br} \& H. Monteiro$^{3}$
\\$^{1}$SUPA, School of Physics \& Astronomy, University of St Andrews, North Haugh, St Andrews, Fife KY16 9SS, UK 
\\$^{2}$Escola de Artes, 
Ci\^encias e Humanidades, 
Universidade de S\~ao Paulo, Rua Arlindo Bettio 1000, CEP 03828-000,
S\~ao Paulo, Brazil
\\$^{3}$Departamento de
  F\'isica, Universidade Federal de Itajub\'a, Av. BPS 1303 -
  Pinheirinho, CEP 37500-903, Itajub\'a, Brazil}

\begin{document}

\date{}

\pagerange{\pageref{firstpage}--\pageref{lastpage}} \pubyear{2012}

\maketitle

\label{firstpage}

\begin{abstract} 
 
  For the past decade observations of the alignement of PNe symmetries
  with respect to the galactic disk have led to conflicting
  results. Recently the first direct observational evidence for a
    real alignment between PNe and local interstellar magnetic fields
    in the central part of the Galaxy ($b < 5^\circ$) has been
    found. Motivated by the recent dicovery we studied the role of the
    interstellar magnetic field on the dynamical evolution of a
    planetary nebula by means of an analytical model and from 3D MHD
    numerical simulations.  In our models the nebula is the result of
    a short-time event of mass ejection with its surrounding
    medium. The nebula asphericity is assumed to be due to an
    intrinsic shaping mechanism, dominated by the latitude-dependent
    AGB wind, and not the ISM field. We test under what conditions
  typical ejecta would have their dynamics severely modified by an
  interstellar magnetic field. We found that uniform fields of $>
  100\mu$G are required in order to be dynamically dominant. This is
  found to occur only at later evolutionary stages, therefore being
  unable to change the general morphology of the nebula. However, the
  symmetry axis of bipolar and elliptical nebulae end up aligned to
  the external field. This result can explain why different
  samples of PNe result in different conclusions regarding the
  alignment of PNe. Objects located at high galactic latitudes, or at
  large radii, should present no preferential alignment with respect
  to the galactic plane. PNe located at the galactic centre and low
  latitudes would, on the other hand, be preferentiably aligned to the
  disk. Finally, we present synthetic polarization maps of the nebulae
  to show that the polarization vectors, as well as the field lines at
  the expanding shell, are not uniform even in the strongly magnetized
  case, indicating that polarization maps of nebulae are not
    adequate in probing the orientation, or intensity, of the
  dominant external field.

\end{abstract}

      
\section{Introduction}

Planetary nebulae are end products of the evolution of stars with
masses below $8 M_{\odot}$. The importance of these objects extends
well beyond the physics that lead to the ejection of the outer layers of stars. 
PNe have also been key on understanding basic atomic processes, providing
powerfull tools to probe the physical and chemical
characteristics of our Galaxy, on the understanding of the internal 
processes of low and intermediate mass stars, and others.

Since the work of \citet{C18}, many attempts have been made to
classify PNe according to their morphology and correlate these with
basic properties such as central star (CS)
temperature, mass, position in the Galaxy, and other parameters
\citep[see
e.g.][]{KK68,B87,BP87,BPI87,B97,IPB89,M90,S93,ZK98,S99,M04}.  The
miryad of different shapes observed can be basically classified in
four major goups: round/spherical, ellipticals, bipolars and irregulars.  It is
not clear yet what process is dominant in shaping each morphology. 
A good summary of how morphological studies have
been employed over the last half-decade to understand shaping
mechanisms is given by \cite{S12}. Possibly more than one can
operate in each object. Among the most plausible processes 
presented so far we have interacting
outflows, magnetic fields, binarity, and even the interaction of the
ejecta with the interstellar medium (ISM).

The role of the ISM in shaping evolved PNe has been sistematically
investigated since the work of \citet{gurzadyan69}. A recent and
extensive study on the classification of interacting PNe is presented
by \cite{ali12}. \citet{wu11} related the morphological features observed 
in several PNe with their relative motion with respect to the ISM.  
The dynamical interaction between the
ISM and the PNe would occur basically in two ways: interaction with
the interstellar magnetic field and due to a relative velocity between
the central source and the surrounding medium. Analytical 
estimates \citep[e.g.][]{smith76} and
numerical simulations \citep[e.g.][]{soker97,war07} predict changes in
nebular shape, as well as anisotropies in density and emission, when
the central source is moving through the ISM. These are obviously
unable to generally explain the different PN morphologies, but are quite
successful in explaining the comet-like nebulae observed, as well as
misplaced central stars with respect to the geometric centre of the
nebula.

Regarding the magnetic fields, theoretical studies have been mostly
focused on the early shaping of the nebulae, right after the 
envelope ejection \citep[e.g.][]{heiligman80,
  pascoli85,stone92,balfrank02,garcia97,garcia99,matt06,black08}.  The
magnetic interaction of ISM magnetic fields and PNe has been just
recently observed in Sh 2-216
\citep{ransom08}. These authors detected Faraday rotation along the
PNe and were able to estimate the magnetic field intensity at the
interaction region as $\sim 5 \pm 2 \mu$G. The asymmetric radio
emission revealed the compression of the ISM field by the
nebula. These observations are in agreement, for instance, with the
models of \citet{soker97}. In this particular case though, there is
little evidence for any role of the magnetic pressure in modifying the
morphology of the nebula.

Apparent alignment of aspherical PNe with the interstellar magnetic
field was first reported by \citet{grinin68}. \citet{melnick75} and
\citet{phillips97} claim to have obtained statistically relevant 
alignment between the major axis of eliptical and bipolar PNe with 
respect to the the galactic plane. \citet{corradi98} on the other hand 
studied a larger sample of objects getting no preferential alignment, 
and concluded that PNe are randomly oriented on the sky. These authors
however did not compare the orientation of the PNe with estimates of
the magnetic field orientation. \citet{grinin68} showed
that the correlation between the nebula axis of symmetry and the
interstellar magnetic field orientation, probed by linear
polarization, is much stronger than the correlation for the galactic
latitude. More recently, \citet{weid08} analysed over 400 objects
all-sky and concluded that, as pointed by \citet{corradi98}, PNe are
in general not preferentially aligned to the galactic plane.
However, these authors found a correlation between the orientation of the major
axis of some PNe and the galactic plane for a small region of the sky,
near the galactic centre.  

 In a recent work, \citet{rees13} provided strong observational
  evidence for the alignment between PNe and the galactic plane. They
  studied over 100 PNe at the galactic centre with positive detection
  of alignment for bipolar nebulae. The statistics presented showed no
  relevant alignment for the other morphological types.  These authors
  suggest that strong magnetic fields acted during the formation of
  the stars, driving a global alignment of stellar angular momenta
  with the large scale magnetic field. 

The considerations above lead naturally to the following questions:

\begin{itemize}
\item is it possible for the interstellar magnetic fields to tilt the 
PNe axis of symmetry, specially when these are initially misaligned 
with respect to the external field?

\item and, why does this effect seem to occur preferably at the
  galactic centre?
\end{itemize}

Despite the achievements of the last few decades, no systematic study
of the dynamical role of interstellar magnetic fields on the later
evolution of PNe, with originally misaligned symmetries with respect
to the ISM magnetic field, have been performed so far.  This is
exactly the problem we address in this paper.

In this work we propose an alternative scenario to that proposed by 
\citet{rees13}. We study the role of the interstellar magnetic field in
changing, or distorting, the PNe large scale morphologies.  We
  assume that the initial shaping of the nebulae is done mainly by
  interacting winds, and magnetic fields have little effect in these
  initial stages. The effects of the external field appear at later
stages, causing the original nebular asymmetry to be modified.  We
study if typical galactic fields are able to distort the PN
morphollogies to account for the observed fraction. The manuscript is
organized as follows.  A simple analytic description of the dynamical
evolution of a blast wave is provided in Section 2. The analytical
estimates are then tested numerically, under a more detailed
multi-dimensional study, as shown in Section 3, followed by the
Conclusions.

\section{Dynamics of the expanding shell}

The expansion of a spherical blast wave has been extensively studied
theoretically in the past, in different contexts from acoustic blast
waves in the atmosphere \citep{taylor50}, stellar winds
\citep{weaver77} to SN remnants \citep{wol72, melioli06}, and in the
the magnetized case \citep[e.g.][]{heiligman80,stone92,soker97,Leao09}.

There is no consensus yet on the detailed physics that triggers the post-AGB 
superwinds. Current stellar evolution models study the role of angular 
momentum, pulsations and magnetic fields on them. Despite model 
uncertainties it is clear that the process of 
evelope ejection is very fast, and must occur at very short 
transition timescales ($\tau_{\rm trans} \ll 10^3$yrs) compared 
to the lifetimes of observed PNe ($t \sim 10^4$yrs) \citep[see][for a 
review]{winc03}. Besides, assuming a typical expansion velocity of 
$\sim 100$km s$^{-1}$, the nebula is mass/energy loaded up to a 
lengthscale $l<10^{-3}$pc. Both time and length scales for the 
superwind to load the nebula are very small compared to those we 
are interested in this work, and we may then assume the energy injection 
as quasi instantaneous. In this case we can model the
expansion of the PN over the ISM as a single blast.

For the sake of simplicity let us consider a planetary nebula as the
result of a total mass ejection $M_0$, occuring in a short time event
(shorter than the typical expansion timescales) with initial energy
$E_0$. After the burst the material adiabatically expands over the
surrounding ambient medium. The expansion occurs initially due to the
work done by the internal pressure over the ambient cold gas, in which
pressure is negligible compared to the total energy density of the
blast wave. The radius of the expanding shell $R_{\rm s}$ is obtained
as a function of time by assuming that energy is conserved as:

\begin{equation}
R_{\rm s}^{\rm adi}(t) \sim \left( \frac{E_0}{\rho_{\rm ISM}} \right)^{1/5} t^{2/5},
\end{equation}

\noindent
where $\rho_{\rm ISM}$ represents the mass density of the ISM. The velocity 
is readily obtained as $V_{\rm s}^{\rm adi}(t) \sim \left( \frac{E_0}{\rho_{\rm ISM}} 
\right)^{1/5} t^{-3/5}$.

The adiabatic phase lasts until the radiative losses become relevant at the shock, i.e. 
when $\int \dot{E}_{\rm rad}dt' \sim E(t)$. The determination of the transition time
($t=\tau_{\rm rad}$), between the adiabatic and radiative phases, depends on the radiative 
cooling curve assumed. For typical solar abundandances \citep{weaver77}:

\begin{equation}
\int \dot{E}_{\rm rad}dt' \simeq 0.32 n_{\rm ISM}^{9/5} E_0^{1/5} t^{17/5},
\end{equation}

\noindent
which gives a transition timescale of:

\begin{equation}
\tau_{\rm rad} \sim E_0^{4/17} n_{\rm ISM}^{-9/5}.
\end{equation}

\noindent
For $E_0 \sim 10^{45}$ergs and $n_{\rm ISM} \sim 1$cm$^{-3}$ one 
obtains $\tau_{\rm rad} \sim 300$yrs, which is short compared 
to the typical PNe lifetimes $\sim 10^4$yrs. If $R_{\rm rad}$ and $V_{\rm s}$ 
correspond to the shell radius and velocity at the transition time
the values given above result in $R_{\rm rad} \sim 0.04$pc and 
$V_{\rm s,rad} \sim 200$km s$^{-1}$.

At $t > \tau_{\rm rad}$ momentum conservation can be used instead, i.e. $R_{\rm s}^3 V_{\rm s} \rho_{\rm ISM} \simeq$ const., which results in:

\begin{equation}
R_{\rm s} \simeq R_{\rm rad} \left( \frac{8}{5} \frac{t}{\tau_{\rm rad}} - \frac{3}{5} \right)^{\frac{1}{4}},
\end{equation}

\noindent
and

\begin{equation}
V_{\rm s} \simeq V_{\rm rad} \left( \frac{8}{5} \frac{t}{\tau_{\rm rad}} - \frac{3}{5} \right)^{-\frac{3}{4}}.
\end{equation}

The total mass of interstellar gas that is swept up and acumulates at the expanding shell is approximately given by $M_{\rm s} \simeq 
\rho_{\rm ISM} R_{\rm s}^3 (4 \pi /3)$. Therefore, the kinetic energy density, i.e. the ram pressure, may be estimated as 

\begin{equation}
p_{\rm ram} \simeq \rho_{\rm ISM} V_{\rm s}^2 \frac{R_{\rm s}^3}{\left[R_{\rm s}^3-(R_{\rm s}-\Delta)^3 \right]}
\end{equation}

being $\Delta$ the thickness of the outer shock region, defined as the region between the contact discontinuity and 
the shock surface between the ISM and the expanding shell. 
Since the radiative losses are assumed to be very efficient $\Delta \ll R_{\rm s}$ and $p_{\rm ram} \sim \rho_{\rm ISM} V_{\rm s}^2 (1-3\Delta)^{-1}$.
The shell ceases its expansion once the ram pressure is reduced, and balanced with the ambient pressure.

\subsection{Magnetic fields}

Once the ISM is considered magnetized not only the ambient gas is dragged and acumulated at the expanding shell, but also the magnetic 
field lines. The result is an increase in the total pressure at the outer shock, and a reduction on the expansion of the nebula 
compared to the pure hydrodynamical case. 

The process is understood as follows. In the hydrodynamical case most of the kinetic 
energy of the shock wave is transfered to the 
ambient gas as thermal energy. The remaining energy is kept as kinetic 
for the shocked gas to keep expanding. In the magneto-hydrodynamical 
(MHD) case, part of the energy of the shock wave that would be transformed 
into thermal energy is actually converted into magnetic one. This occurs 
once the magnetic field lines are compressed together with the ambient gas. 
The magnetic pressure increases in the outer shock region and, as 
a consequence, the temperature of the shock gas is lower compared to 
the hydrodynamical case. 
The magnetic pressure acts in the direction perpendicular to the field 
lines, i.e. for ${\bf V}_{\rm s} \perp {\bf B_0}$. There is 
free expansion along the field lines. 

Let us reconsider the problem of a spherical blast wave propagating over a homogeneous ambient medium, 
but now threaded by an uniform magnetic field $B_0$. Also, let us consider that the ISM magnetic 
field is weak compared to the kinetic energy density of the blast. This assumption allows us to treat the 
expansion as isotropic and the results of Eqs.\ 3-5 remain valid. 
Once the shell reaches a radius $R_{\rm s}$, at $t>\tau_{\rm rad}$, 
it has also dragged interstellar magnetic field lines into the outer shock layer. 
The magnetic pressure is given by:

\begin{equation}
p_{\rm mag}(\theta) \simeq \frac{B_0^2}{8\pi} \left| \sin{\theta} \right| \frac{R_{\rm s}^3}{\left[R_{\rm s}^3-(R_{\rm s}-\Delta)^3 \right]},
\end{equation}

\noindent
where $\theta$ is the angle between ${\bf V}_{\rm s}$ and ${\bf B_0}$, at a 
given position over the shell. 
The magnetic pressure is therefore not isotropic at the outer shock region. 
The magnetic energy is acumulated most
at the regions where expansion occured perpendicular to the ambient field lines. 

The free expansion of the nebula will cease at different timescaling depending 
on $\theta$. The expansion will occur as described for the unmagnetized case 
on the direction parallel to the field lines. On the perpendicular direction 
the stall occurs earlier, once $p_{\rm ram} \sim p_{\rm mag}(\theta=\pi/2)$. 

The ratio between the kinetic and magnetic pressure within the outer shock region, 
for $t \gg \tau_{\rm rad}$, is:

\begin{equation}
\beta \equiv \frac{p_{\rm ram}}{p_{\rm mag}(\theta=\pi/2)} \approx \frac{8 \pi \rho_{\rm ISM} V_{\rm rad}^2}{\sin{\theta} B_0^2} \left( \frac{8}{5}\frac{t}{\tau_{\rm rad}} \right)^{-5/4}.
\end{equation}

Initially, considering typical ISM and PN properties, one expects $\beta (\theta) \gg 1$ everywhere in the nebula. 
The transition from kinetic to magnetically dominated dynamics (equipartition) occurs when $\beta \simeq 1$, i.e. when:

\begin{eqnarray}
\tau_{\rm equip} \approx 2.5 \times 10^{-2} \left(\frac{E_0}{10^{45} {\rm ergs}} \right)^{4/17} \left(\frac{V_{\rm s,rad}}{200 {\rm km\ s^{-1}}} \right)^{8/5} \times \nonumber \\ \left(\frac{n_{\rm ISM}}{1 {\rm cm^{-3}}} \right)^{23/85}\sin{\theta}^{-4/5} B_0^{-8/5} \ {\rm yrs}.
\end{eqnarray}

\begin{figure}
 \includegraphics[width=8.3cm]{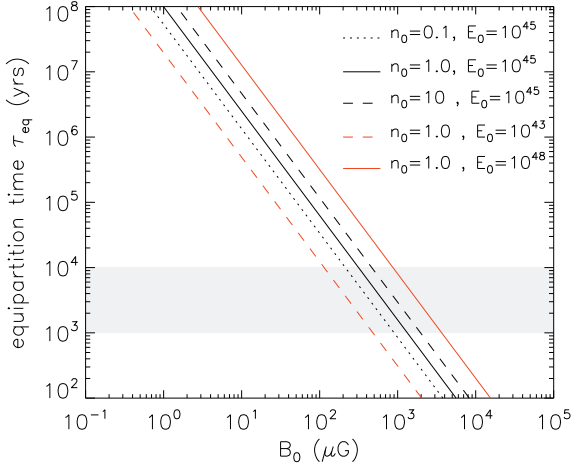}
 \caption{The equipartition timescale $\tau_{\rm equip}$ as a function of the ambient magnetic field $B_0$. The lines correspond to different 
 sets of initial energy input $E_0$ and interstellar density $n_{\rm ISM}$. Black lines were obtained for $E_0=10^{45}$ergs 
 with $n_{\rm ISM}=0.1$cm$^{-3}$ (dotted), $1$cm$^{-3}$ (solid) and $10$cm$^{-3}$ (dashed). Red lines were obtained for $n_{\rm ISM}=0.1$cm$^{-3}$ 
 with $E_0=10^{48}$ergs (solid) and $10^{42}$ergs (dashed). The grey area 
 corresponds to $\tau_{\rm equip} < 10^4$yrs, i.e. shorter than the typical 
 ages of planetary nebulae.}
 \label{equip}
\end{figure}

From the equation above, for a typical local ISM magnetic field of 
$B_0 = 2-5\mu$G one obtains $\tau_{\rm equip} \simeq 10^7$yrs, 
for $\theta = \pi/2$. This timescale is too large compared to the 
typical ages of PNe, $\sim 10^4$yrs, and, in this case, 
the magnetic fields would have little effect on the dynamics of the 
nebula. The timescale $\tau_{\rm equip}$ is shown in Figure 1 as 
a function of $B_0$ for different interstellar gas densities. The 
grey area corresponds to the typical observed range of ages. We infer 
from this simple model that the dynamics of a PN will be modified 
by the ISM magnetic fields for $B_0 > 100\mu$G. 

Notice that for the estimate above we have disregarded the stellar 
magnetic field. The reason is that surface magnetic fields in the 
range of $B_* \sim 10^{-3} - 10^{-2}$G result in magnetic pressures of $10^{-8} - 10^{-6}$ erg cm$^{-3}$, 
many orders of magnitude smaller than the ram pressure of the wind ($>10^{-2}$ erg cm$^{-3}$). 
If we consider the magnetic field intensity to decay with $r^{-2}$, as for a monopolar configuration 
in a super-Alfv\'enic wind, the influence of the magnetic field of the star is even weaker at larger distances. Except for extreme cases (with $B_* \gg 1$G), 
the stellar magnetic field is irrelevant for the late dynamics of the nebula, and may be neglected.

Obviously, the magnetized case is 
essentially anisotropic and a more detailed and multi-dimensional
analysis is needed for a complete understanding of the problem. 
Also, the influence of the magnetic field during the adiabatic phase 
was not taken into account, as well as other features such as the 
generation and propagation of magnetosonic and Alfv\'en waves. 
In this sense we provide a numerical analysis of the problem in the 
following section with the main goal of testing the validity of 
the previous estimates.

\section{MHD numerical simulations of PN ejecta}

During their asymptotic giant branch (AGB and pre-PN phases) intermediate mass stars are 
supposed to preset strong winds. Depending of stellar rotation or other 
mechanisms, such as binarity or large scale magnetic fields, the 
winds may present latitudinal dependence, 
which result in a circunstellar matter density distribution that is also 
aspherical.
The terminal phases of envelope ejection, at the end of the pre-PN phase, is expected to 
interact with the anisotropic medium that will change its expansion velocities. 
The result is a global morphology that can vary from a spherical shell, to elliptical and 
bipolar ones. The so-called ``interacting winds'' model has been quite successful in 
explaining global morphologies of PNe \citep{IBF92}. 

A proper physical description of the formation of a PN would 
be the result of a complete modelling of the stellar winds during the 
AGB and post-AGB phases, together with a comprehensive study of the
internal conditions during the formation of the white dwarf and how it 
decouples from the envelope. Such details may be important but cannot be 
treated properly in our models. We must simplify the initial 
setup of the problem and focus simply on the dynamical evolution of the 
nebula as it iteracts with the surrounding magnetized ISM.

AGB stars present massive stellar winds, with mass loss rates of
$\dot{M} \sim 10^{-9} - 10^{-5}$M$_{\odot}$yr$^{-1}$, and terminal
velocities of $u_{\infty} \sim 10 - 100$km s$^{-1}$, for periods as
long as 100,000 yrs \citep{fal02,vid06,fal06}. Considering a typical low density ($n \leq 1$
cm$^{-3}$) and weakly magnetized ISM ($B \sim 1 - 5\mu$G), such a wind
would carve the interstellar gas to create an asteropause as large as
$\sim 0.1 - 1$pc. Large ISM turbulent/thermal pressure, or strong
magnetic fields, would naturally reduce this estimate. In the case of
$B > 100\mu$G, as discussed in the previous section, the asteropause
would be limited to radii $< 0.01$pc. The wind-ISM coupling results in
two distinct regions for the expanding PN. One is defined from the
star up to few thousands of AUs, where the ISM influence is
negligible, and the stellar ejecta interact only with the extended
envelope, i.e. the previously ejected stellar winds.  The second
region is defined above the mentioned region, from which the ISM
magnetic pressure dominates the dynamics of the nebula.  From this
scenario it is straightforward to assume the general morphology of the
PNe (round, elliptical or bipolar) to be determined by the interaction
of the post-AGB superwind with the earlier aspherical AGB wind, with
little effect of the ISM magnetic field.  However, once the nebula
reaches the asteropause it should interact with an unperturbed
external ISM.

As mentioned earlier, there is no consensus about what mechanism 
dominates the shaping of PNe. For this work in particular, 
the actual shaping mechanism is irrelevant since we focus on the interaction of 
an already shaped nebula with the surrounding ISM. A straightfoward 
way of generating axisymmetric nebulae for our models is to use the 
interacting winds method. The post-AGB superwind is driven and expands 
over a preset gas distribution which is latitude dependent. The enhanced 
density at the equatorial plane results in slower expansion velocities on 
those directions. The end product is a latitude-dependent expanding 
nebula. Few prescriptions for the preset density distribution were given \citep[e.g.][]{kahn85,IPB89,mel91},
though none of these based on self-consistently driven wind models. For the 
sake of simplicity we use the same setup as in \citet{monteiro11}. 
In such model, the preset ambient gas 
distribution - determined during the stellar AGB phase - should 
follow the distribution \citep[as used by][]{IPB89,mel91}:

\begin{equation}
\rho_{\rm amb}=\frac{\rho_0}{A(\theta)}\left( \frac{r_0}{r} \right) ^2 ,
\end{equation}

\noindent
where

\begin{equation}
A(\theta) = 1-\alpha \left( \frac{e^{\delta cos 2 \theta - \delta}-1}{e^{2 \delta}-1} 
\right),
\end{equation}

being the parameter $\alpha$ is related to the density ratio at the polar
and equatorial directions, and $\delta$ the steepness of the density
profile with the latitude. 

The above equations represent an initial setup for the matter distribution, together 
with the implementation of an energy source (blast wave) at the stellar location, and the 
external mgnetic field. The interaction of the blast wave with the preset density distributuion, 
an later with the external magnetic field, will determine the final global morphology of 
the nebula.

\subsection{Governing equations and numerical setup}

The dynamical evolution of the blast wave, and of the magnetized 
interstellar medium, is governed by the magnetohydrodynamic 
equations (MHD), which can be written in the conservative form:

\begin{equation}
\partial_t\mbox{\bf U} + \nabla \cdot\mbox{\bf F}(\mbox{\bf U}) = f(\mbox{\bf U}),
\label{eq:genform}
\end{equation}

\noindent
where $f({\bf U})$ is the source term, ${\bf U}$ is the vector of conserved variables:

\begin{equation}
{\bf U} = \left( \rho, \rho{\bf v}, \left(\frac{1}{\gamma-1} p + \frac{1}{2}\rho v^2 + \frac{B^2}{2}\right), {\bf B} \right)^T,
\end{equation}

\noindent
and ${\bf F}$ is the flux tensor:

\begin{eqnarray}
{\bf F} & = & \Bigl( \rho {\bf v}, \rho {\bf v}{\bf v}+ p_{\rm tot}{\bf I} - {\bf B}{\bf B}, \left(\frac{\gamma}{\gamma -1}p + \frac{1}{2}\rho v^2 \right){\bf v} \nonumber \\
        &   &  -{\bf B}\left({\bf B}{\bf v} \right), {\bf v} {\bf B} - {\bf B} {\bf v}\Bigr)^{T}
\label{eq:flux}
\end{eqnarray}

\noindent
where $\rho$ is the gas mass density, ${\bf v}$ the fluid velocity,
${\bf B}$ the magnetic field, $p$ the thermal pressure, $p_{\rm tot}=p+B^2/8\pi$
$p_{tot} = p + p_{\rm mag}$ the total pressure, and 
$\gamma$ the adiabatic polytropic index, and $f$ corresponds to 
source terms for the given conserved variable $U$. The set of equations is 
closed by calculating the radiative cooling as source term for the energy 
equation, as follows:

\begin{equation}
\frac{\partial p}{\partial t} = \frac{1}{(1-\gamma)} n^2 \Lambda(T),
\end{equation}

\noindent
where $n$ is the number density and $\Lambda(T)$ is the cooling function, which is 
obtained through an interpolation method of the electron cooling efficiency 
table for an optically thin gas \citep{gnat07}.

In the simulations the above set of equations was solved using the GODUNOV 
code\footnote{http://amuncode.org}, which has been extensively tested 
and used in many astrophysical problems \citep[e.g.][]{kowal10, fal10a, 
fal10b, fal10c, kowal11a, kowal11b, fal11, kowal12, ruiz13, poidevin13}.
 The spatial reconstruction is obtained by a
$5^{th}$ order monotonicity-preserving (MP) method \citep{he11}, with 
approximate HLLC Riemann solver \citep{mignone06}. The time 
integration is performed with the use of a $3^{rd}$ order
four-stage explicit optimal Strong Stability Preserving Runge-Kutta SSPRK(4,3)
method \citep{ruuth06}. For the magnetic field 
we make use of a hyperbolic divergence cleaning approach \citep{dedner02}.

\begin{table*}
\centering
\caption{Parameters used in each simulation.}
\begin{tabular}{ccccccc}
\hline
$Model$ &	$\alpha$ & 	global morphology &	$B_{0}$($\mu$G)	& $\theta$ & dimensions & resolution (pixels) \\
\hline
\hline
SPH1  &	0.0 & 	spherical &	5   &	45$^{\circ}$ & 2.5D & $2048 \times 1024$ \\
SPH2  &	0.0 & 	spherical &	50  &	45$^{\circ}$ & 2.5D & $2048 \times 1024$ \\
SPH3  &	0.0 & 	spherical &	500 &	45$^{\circ}$ & 2.5D & $2048 \times 1024$ \\
SPH4  &	0.20& 	spherical &	5   &	45$^{\circ}$ & 2.5D & $2048 \times 1024$ \\
SPH5  &	0.20& 	spherical &	50  &	45$^{\circ}$ & 2.5D & $2048 \times 1024$ \\
SPH6  &	0.20& 	spherical &	500 &	45$^{\circ}$ & 2.5D & $2048 \times 1024$ \\
SPH7  &	0.20& 	spherical &	5   &	45$^{\circ}$ & 3D 	& $512 \times 256 \times 256$ \\
SPH8  &	0.20& 	spherical &	500 &	45$^{\circ}$ & 3D 	& $512 \times 256 \times 256$ \\
ELI1  &	0.60& 	eliptical &	5   &	45$^{\circ}$ & 2.5D & $2048 \times 1024$ \\
ELI2  &	0.60& 	eliptical &	50  &	45$^{\circ}$ & 2.5D & $2048 \times 1024$ \\
ELI3  &	0.60& 	eliptical &	500 &	45$^{\circ}$ & 2.5D & $2048 \times 1024$ \\
ELI4  &	0.60& 	eliptical &	5   &	45$^{\circ}$ & 3D 	& $512 \times 256 \times 256$ \\
ELI5  &	0.60& 	eliptical &	500 &	45$^{\circ}$ & 3D 	& $512 \times 256 \times 256$ \\
BIP1  &	0.80& 	bipolar   &	5   &	45$^{\circ}$ & 2.5D & $2048 \times 1024$ \\
BIP2  &	0.80& 	bipolar   &	50  &	45$^{\circ}$ & 2.5D & $2048 \times 1024$ \\
BIP3  &	0.80& 	bipolar   &	500 &	45$^{\circ}$ & 2.5D & $2048 \times 1024$ \\
BIP4  &	0.95& 	bipolar   &	5   &	45$^{\circ}$ & 2.5D & $2048 \times 1024$ \\
BIP5  &	0.95& 	bipolar   &	50  &	45$^{\circ}$ & 2.5D & $2048 \times 1024$ \\
BIP6  &	0.95& 	bipolar   &	500 &	45$^{\circ}$ & 2.5D & $2048 \times 1024$ \\
BIP7  &	0.95& 	bipolar   &	500 &	30$^{\circ}$ & 2.5D & $2048 \times 1024$ \\
BIP8  &	0.95& 	bipolar   &	500 &	60$^{\circ}$ & 2.5D & $2048 \times 1024$ \\
BIP9  &	0.95& 	bipolar   &	5   &	45$^{\circ}$ & 3D 	& $512 \times 256 \times 256$ \\
BIP10 &	0.95& 	bipolar   &	500 &	45$^{\circ}$ & 3D 	& $512 \times 256 \times 256$ \\
\hline
\end{tabular}
\label{tab}
\end{table*}

Initially we performed a number of 2.5-dimensional models with high resolution in 
order to obtain the general trends in the observed morphology. 
The computational domain is defined as an uniformly distributed grid, 
in cartesian coordinates, with 2048$\times$1024 cells along $x$ and $y$ 
directions, respectively, corresponding to sizes of 1 and 0.5 in code units. 
Each code unit in length corresponds to 0.2pc, which results in a 
spatial resolution of approximately $9.6 \times 10^{-5}$pc per pixel. Following, 
we selected few models, as decribed in Table\ref{tab}, to run in full 
3-dimensional simulations. For these, due to computational resources, the 
numerical resolution is set as 512$\times$256$\times$256 cells along 
$x$, $y$ and $z$ axis.

The ambient density is initially set as in Eqs.\ 10 and 11. For the sake of 
simplicity in running the models we fixed $\delta = 1.0$ since its role on the global shape of the 
nebulae is less than that of $\alpha = 1.0$.
The blast is initiated by a total energy injection of $10^{45}$ergs 
inside a sphere of radius $r_0 = 0.02$c.u., as in the simulations 
perfomed by \citep{monteiro11}. 
The ambient medium magnetic field is assumed to be initially uniform, set as ${\bf B_0}$, 
with a given angle $\theta$ with respect to the axis of symmetry of the nebula. 
The simulations are performed for a range of intensities $B_0$ and inclinations $\theta$ 
of the ambient magnetic field with respect to the symmetry axis of the nebula.

\subsection{PN morphology and dynamics}

As a general result from the simulations we found that the global morphologies 
obtained are basically independent on the magnetic field intensity, 
at least for the values of $B_0$ set in the simulations. It is clear from Figure\ref{equip} 
that the external magnetic field would only modifify the global morphology of the 
blast wave for $B_0 > 1$mG. Therefore the main parameter for determining the shape of 
the nebula is $\alpha$.

\begin{figure}
{\centering
 \includegraphics[width=8.3cm]{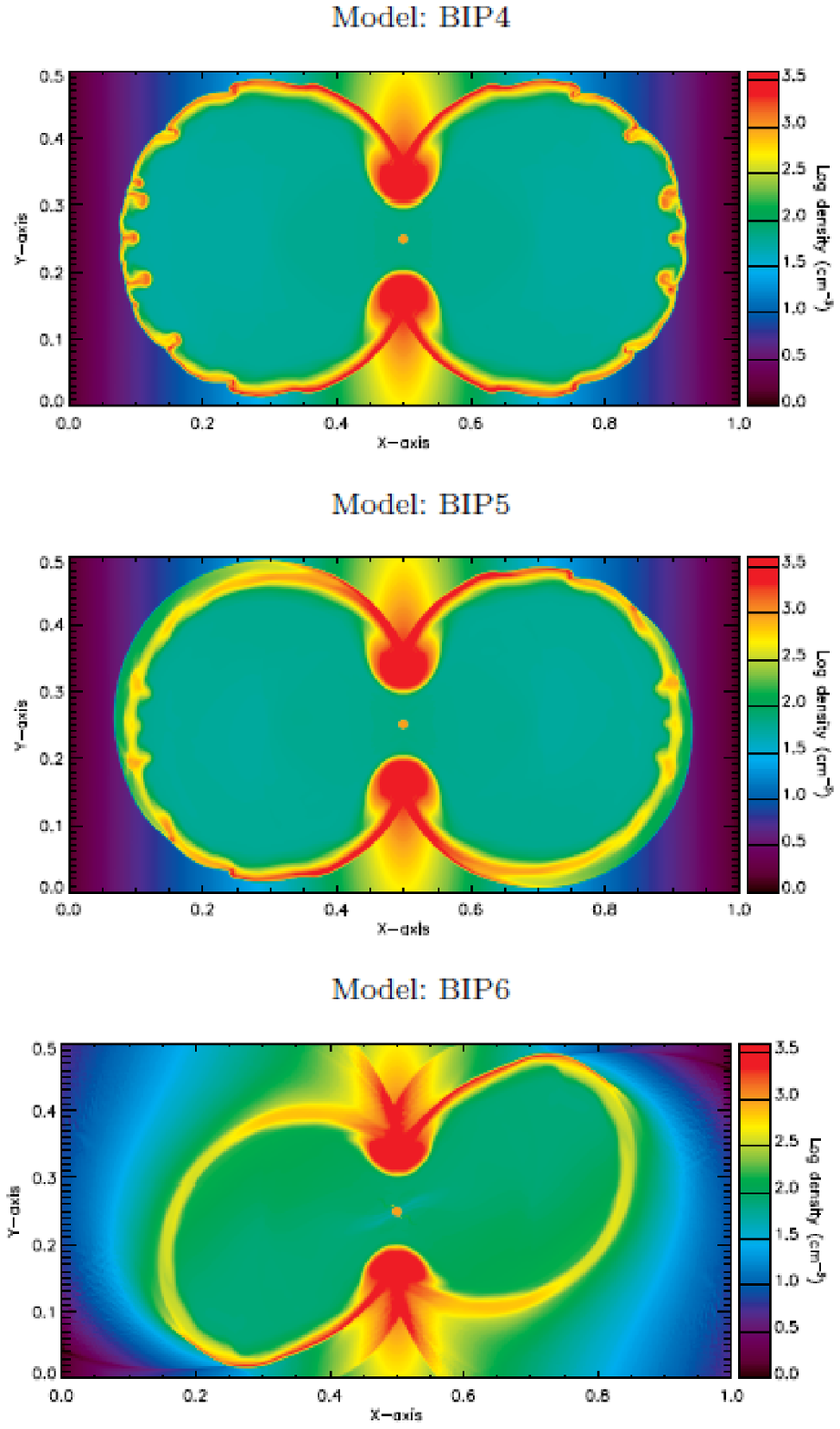}
}
 \caption{Density maps in logarithmic scale for models with $\alpha=0.95$ 
 and $\delta = 0.1$. The ambient magnetic field intensity is initially 
 uniformly set, from top to bottom, as $B_0 = 5$, $50$ and $500\mu$G, respectively. }
\label{densitymaps}
\end{figure}

The main role of the magnetic field in our models is to 
change the expansion velocity of the blast wave at a larger distance from the source. 
In this sense the morphology is not greatly changed - as we show below - but 
the axis of symmetry of the nebula is tilted, depending on the magnetic field intensity. 
We start the morphological and kinematic analysis using the bipolar models as 
basic reference, since these have been used recently by \citet{rees13} to probe the magnetic field 
in the galactic centre.
In Figure\ref{densitymaps} we show the density distributions of models BIP4, BIP5 and BIP6. 
In all three cases the general morphology observed is clearly bipolar with little 
changes in the nebular shapes for increasing magnetic field intensity, except for the 
slightly narrower lobes found for $B_0=500\mu$G. The main difference occurs on the 
orientation of the lobes. For $B_0 = 5\mu$G the axis of symmetry is kept exactly as 
initially set by the density distribution. Small differences are perceived for 
$B_0 = 50\mu$G, where the nebula present an anisotropy in the density distribution. 
However its axis of symmetry is not changed. For $B_0=500\mu$G the density distribution 
of the nebula is greatly changed, with reduced importance of the local instabilities that 
create the knots and clumps. The magnetic tension plays a role as stabilizing source. 
At the same time, the expansion is prevented at the directions perpendicular to the field 
lines. The result is the tilted axis of symmetry of the nebula, in rough agreement with 
the orientation of the external field, i.e. 45$^{\circ}$ with respect to the 
initial axis of symmetry of the preset ambient density distribution.

\begin{figure}
{\centering

 \includegraphics[width=8.5cm]{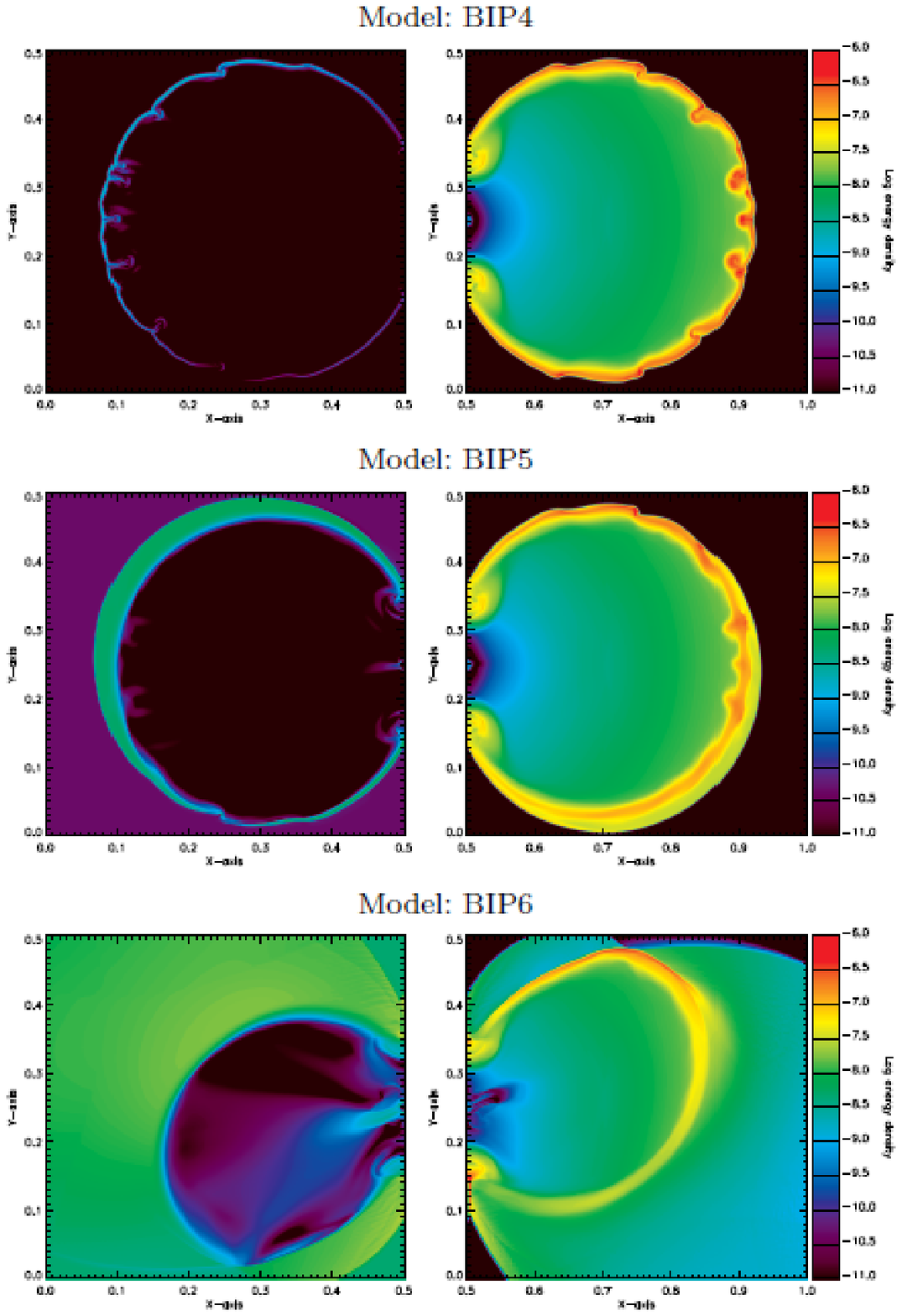}}
  \caption{Magnetic energy density map (left) and kinetic energy 
  density map (right) for models with $\alpha=0.95$ and $\delta = 0.1$. 
  Both maps are in logarithmic scale. 
  The ambient magnetic field intensity is initially uniformly set, 
  from top to bottom, as $B_0 = 5$, $50$ and $500\mu$G, respectively.}
\label{magener}
\end{figure}

In agreement with the dynamical evolution described from an analytical point of view in Section 2, the 
nebula will be tilted by the magnetic field if the kinetic pressure of shell is small, compared to 
the magnetic energy density. This behavior is illustraded quantitatively in Figure \ref{magener} where 
the kinetic and magnetic energy densities are shown for the bipolar nebulae models BIP4, BIP5 and BIP6. 
These maps present the same color table for both physical quantities, and it is possible to directly 
compare the colors in both maps in order to visualize where equipartition is obtained. For $B_0 = 5\mu$G (top) the magnetic energy density (left), $\epsilon_B \sim 10^{-10}-10^{-9}$erg cm$^{-3}$, is clearly 
smaller than the kinetic energy density (right), $\epsilon_{\rm kin} \sim 10^{-7}-10^{-6}$erg cm$^{-3}$. 
The magnetic field intensity is enhanced at the shell though, forming a narrow 
magnetized region due to the pile-up effect that occurs as the nebula drags the ambient field 
lines that acumulate as it expands.

The local increase of the magnetic pressure is more clearly seen 
in the model with $B_0 = 50\mu$G (middle). Here, two different regions are seen in the nebula. One, denser, 
is dominated by the kinetic energy and is less affected by the external field. Surrounding this dense 
shell there is a smooth region where both energies are closely in equipartition. This region is broadened 
due to larger velocities of the perturbations - understood here as Alfv\'en and fast magnetosonic modes -, generated by the compression of the ambient gas. Notice that this broder region is asymmetric with 
respect to the axis of symmetry of the nebula, being broader in the direction perpendicular to the 
external magnetic field. This effect does not occur for model BIP4 because the expansion velocity 
of the shell is larger than those of the Alfv\'en modes. 

Finally, for $B_0 = 500\mu$G (bottom), the 
magnetic energy density is large during most of the simulation and the expansion of the nebula is 
possible to occur on the direction parallel to the external field lines (notably the only regions 
where $\epsilon_{\rm kin}$ is substantially larger than $\epsilon_B$). It is interesting to note here 
that the blast generates magnetosonic waves that propagate fast on the opposite direction, i.e. 
perpendicular to the ambient magnetic field. Such waves have little effect in generating density 
enhancements but could be identified as source of local turbulence, or heating, as they nonlinearly decay.

\begin{figure*}
{\centering
{Model: SPH4	\hspace{20 mm}	Model: SPH6}

 \includegraphics[width=8.3cm]{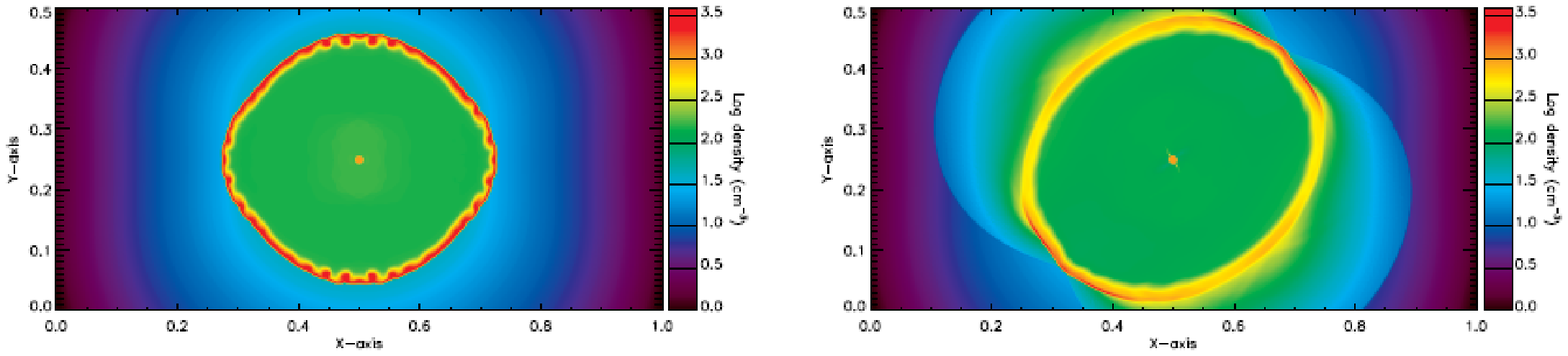} 
 
{Model: ELI1	\hspace{20 mm}	Model: ELI3}

 \includegraphics[width=8.3cm]{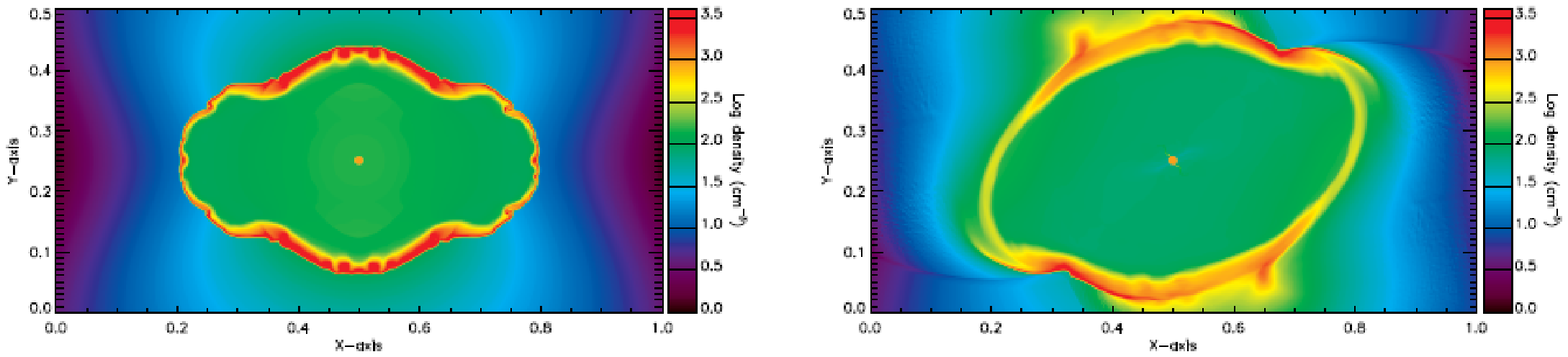} 
 
{Model: BIP1	\hspace{20 mm}	Model: BIP3}

 \includegraphics[width=8.3cm]{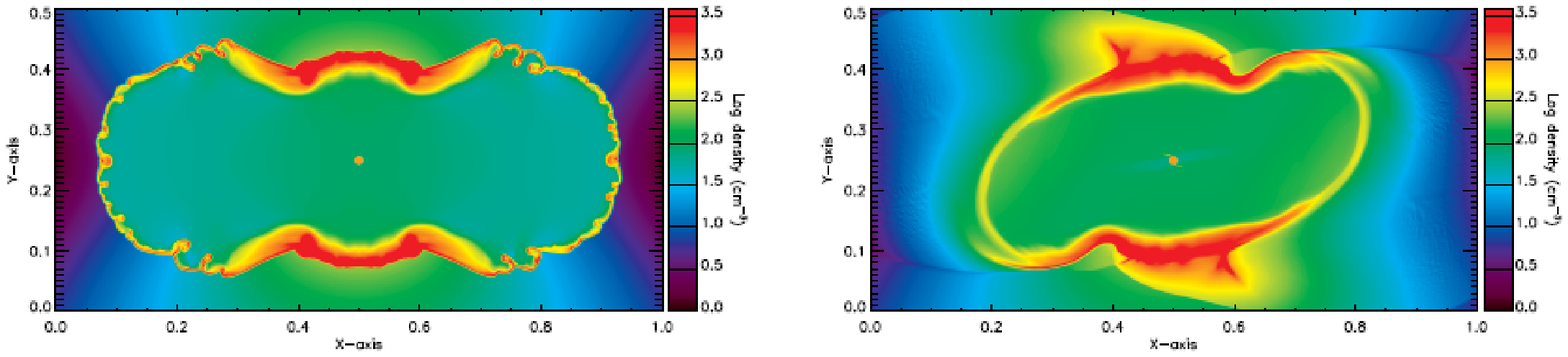}

{Model: BIP4	\hspace{20 mm}	Model: BIP6}

 \includegraphics[width=8.3cm]{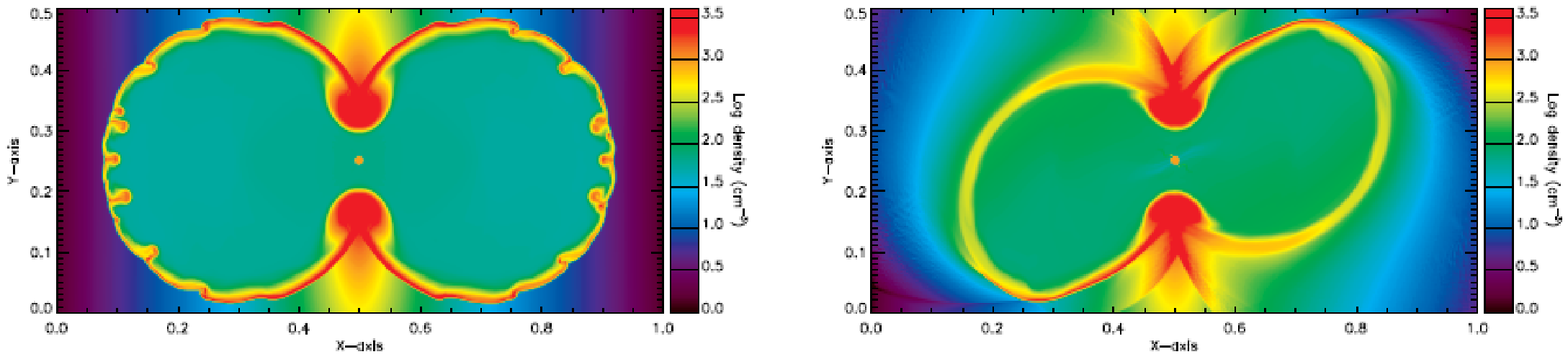}
}
  \caption{Density maps in logarithmic scale for models with, from top to bottom, $\alpha=0.2$, 0.6, 0.8 and 0.95. The ambient magnetic field intensity is initially uniformly set as $B_0 = 5\mu$G (left) and $500\mu$G (right).}
\label{difmodels}
\end{figure*}

All other models run present the same trend as has been discussed above, i.e. all models with $B_0 = 5\mu$G 
showed no changes in morphology due to the external magnetic fields, while the models with 
$B_0 = 500\mu$G presented tilted axis of symmetries aligned to the field lines. We chose 
models that represent each of the morphological groups to be shown in Figure \ref{difmodels}. All models, except SPH6, which is supposed to be spherical, present asymmetries aligned to the 
external field for $B_0=500\mu$G.

\begin{figure*}
{\centering
{Model: BIP9}

 \includegraphics[width=8.0cm]{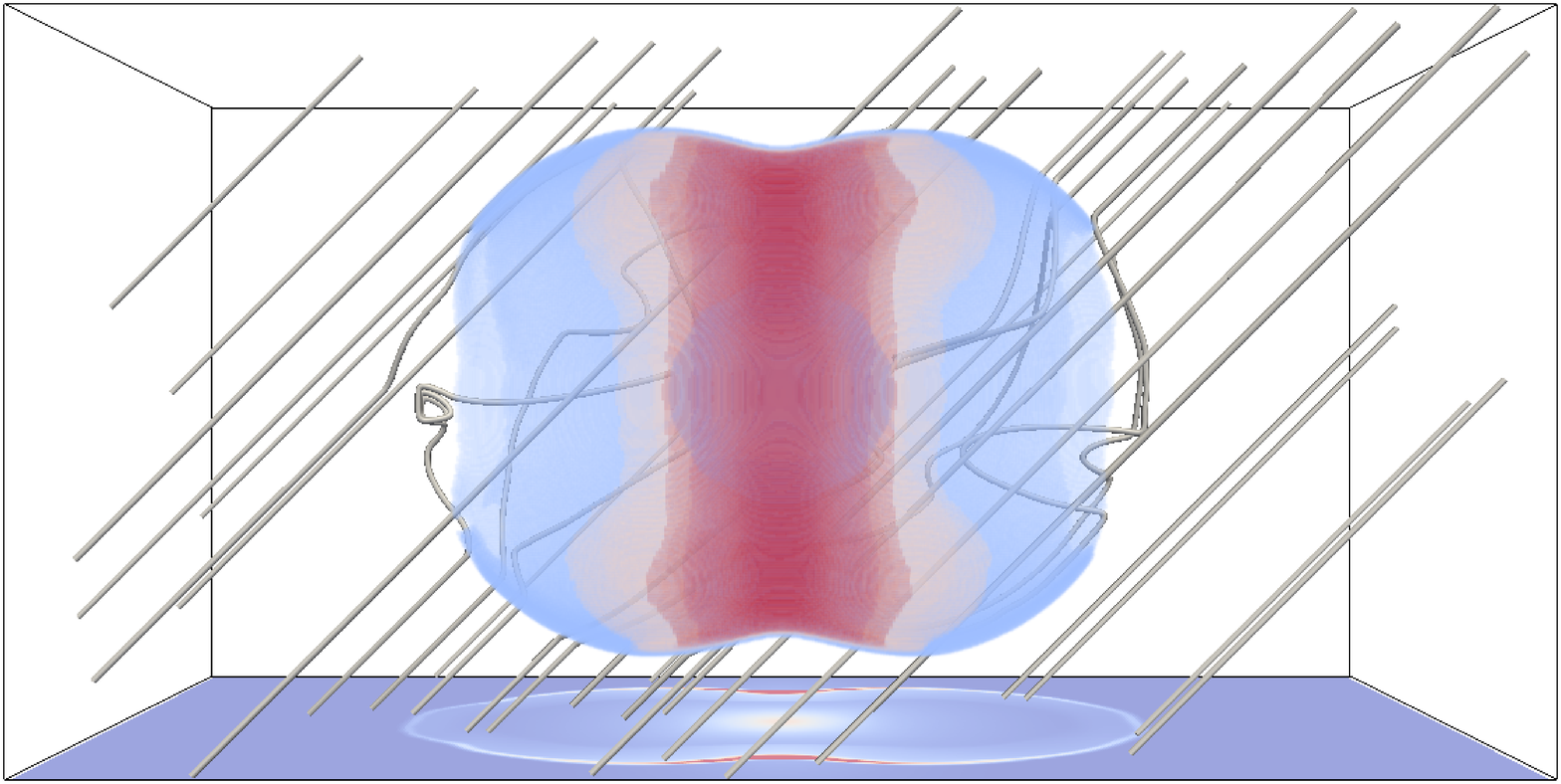}
 \includegraphics[width=7.0cm]{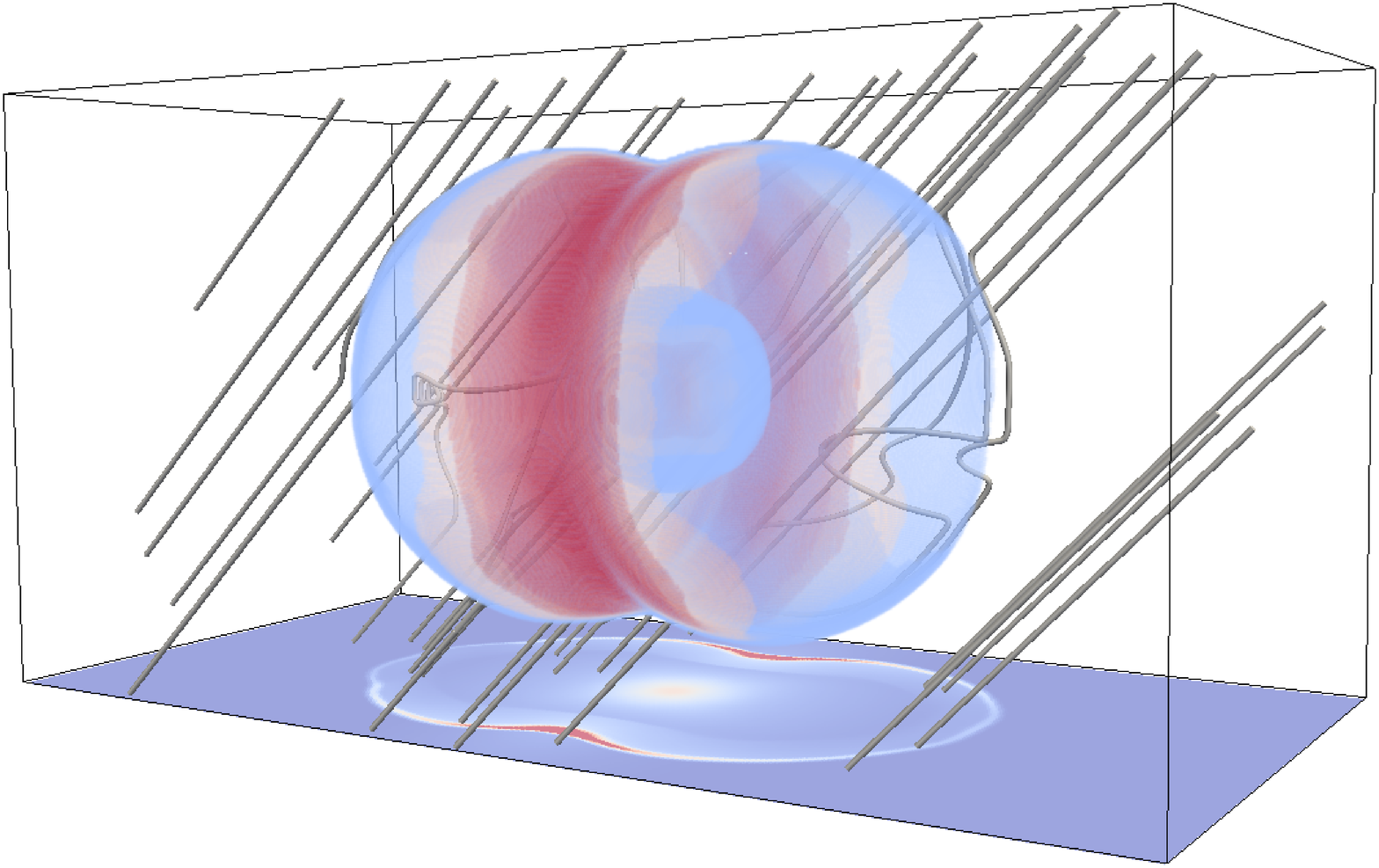} 
 
{Model: BIP10}
  
 \includegraphics[width=8.0cm]{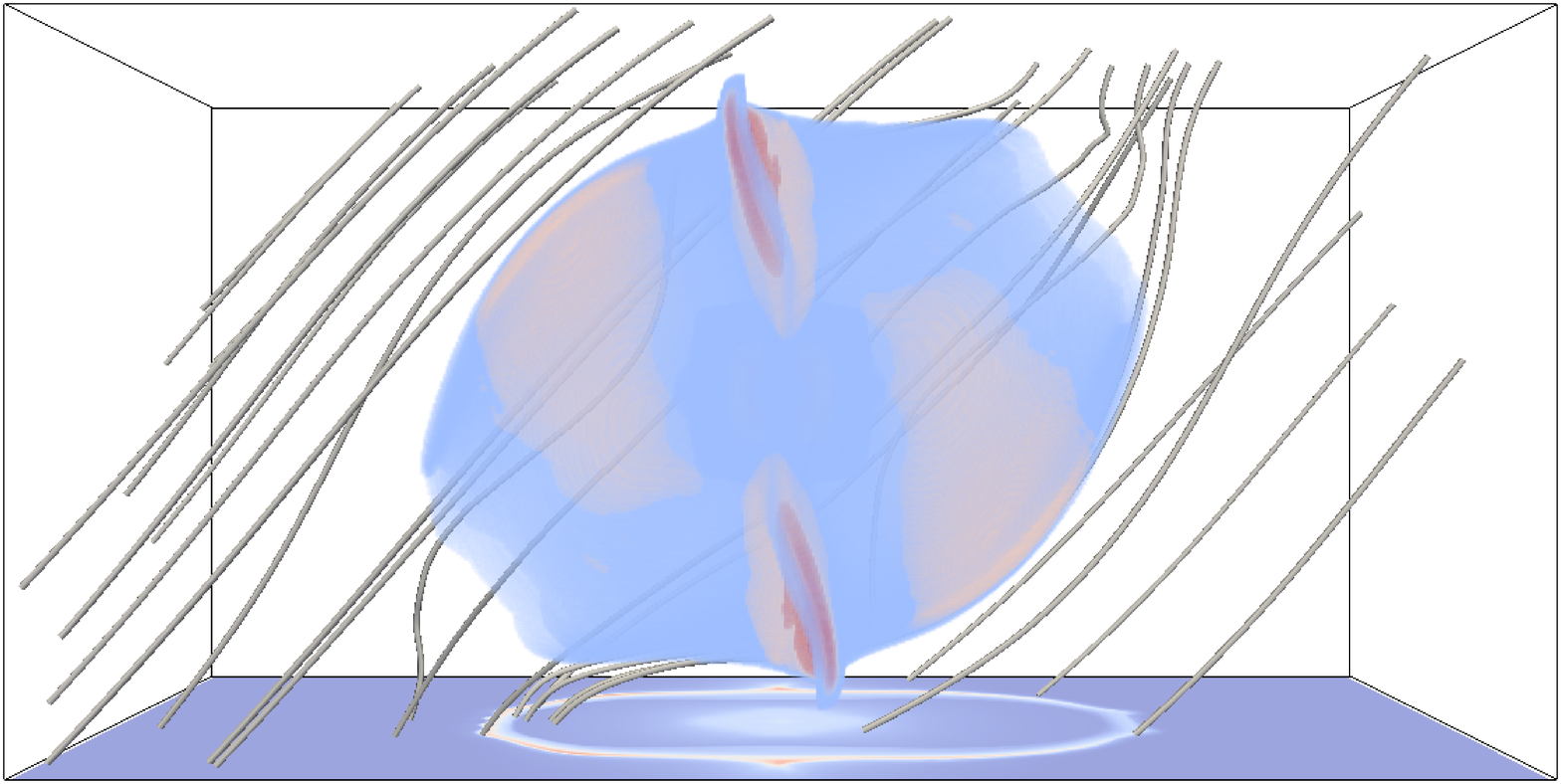}
 \includegraphics[width=7.0cm]{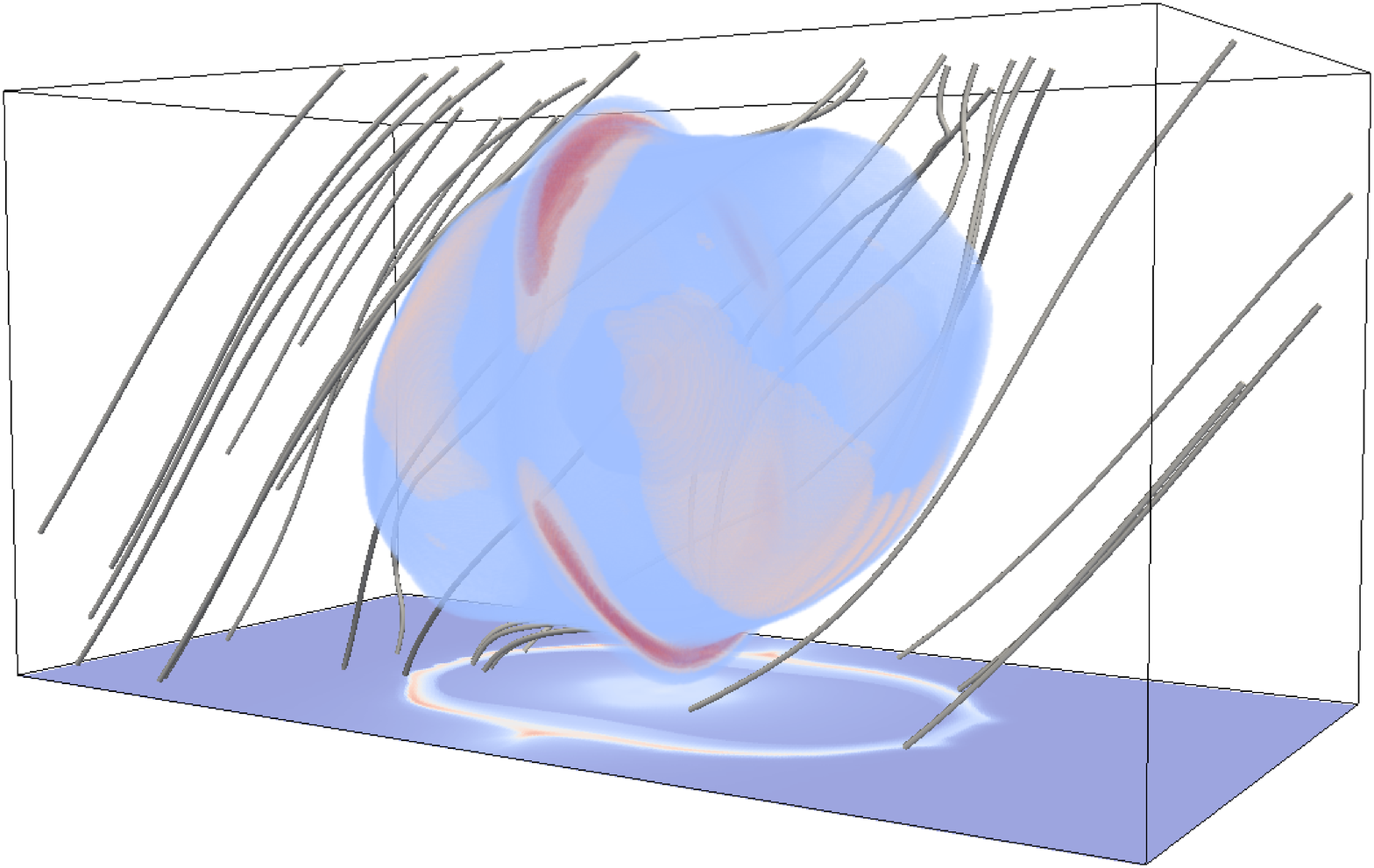} 
 
{Model: ELI5} 
 
 \includegraphics[width=8.0cm]{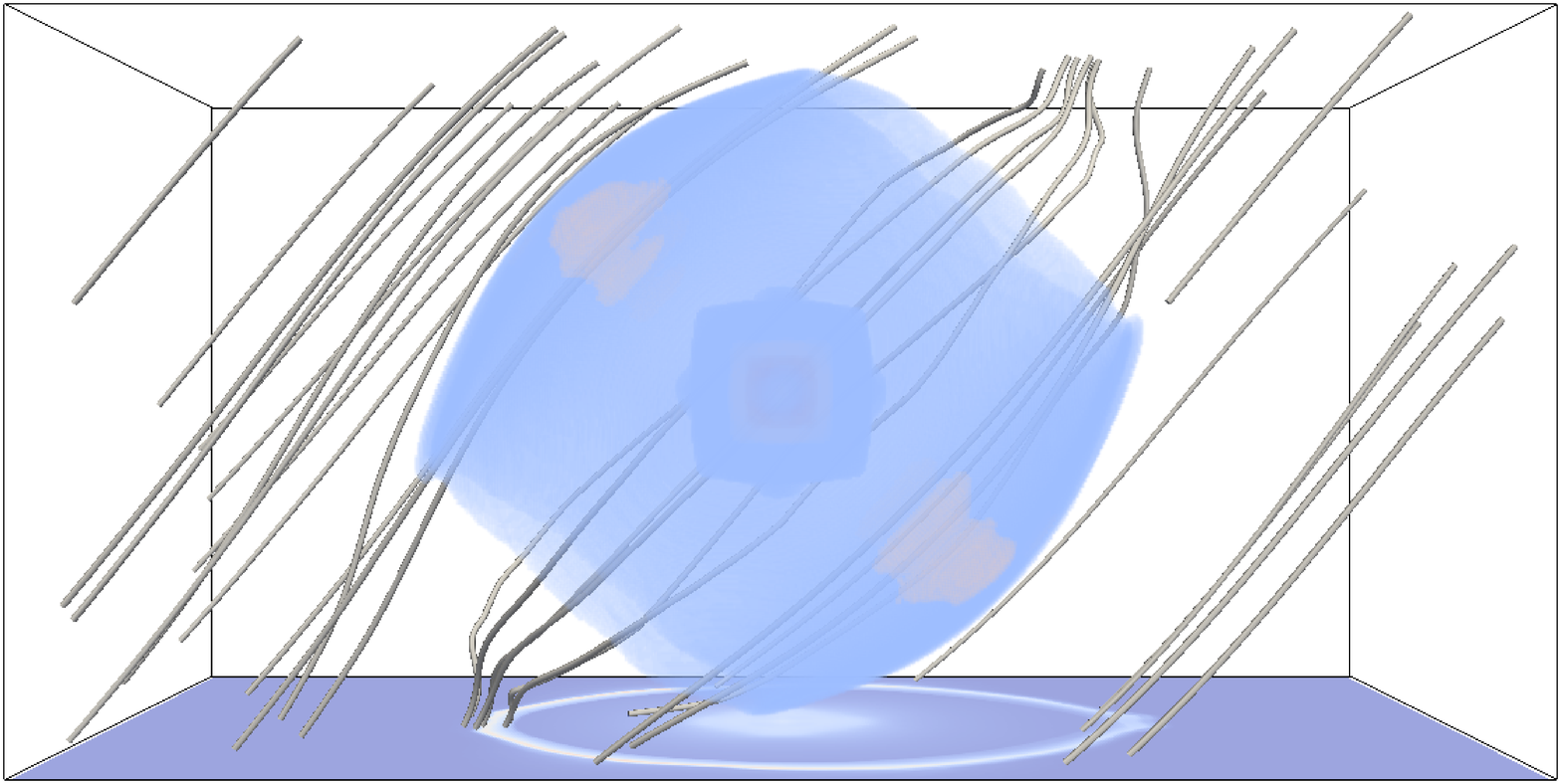}
 \includegraphics[width=7.0cm]{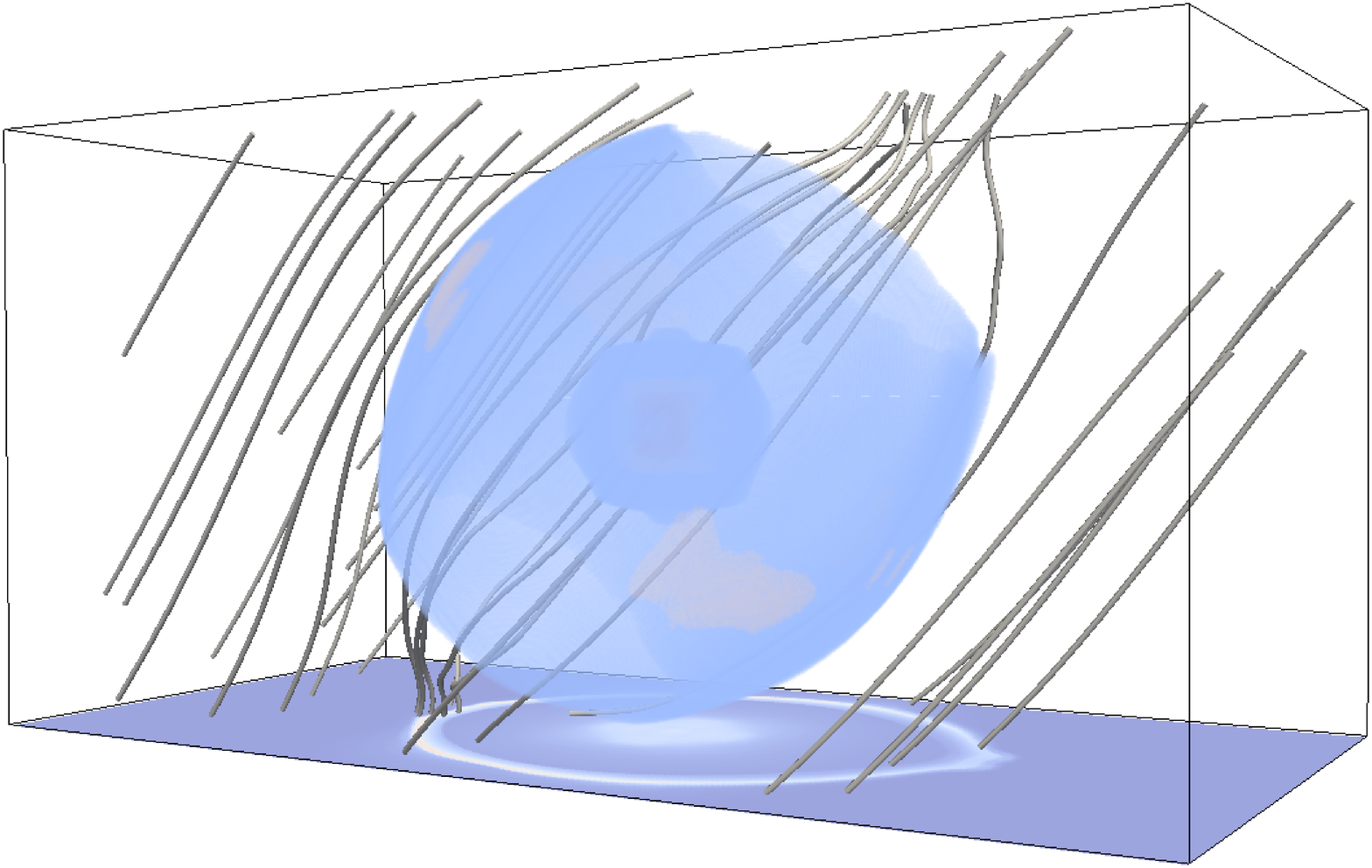} }
  \caption{3-dimensional vizualization of the shell morphology and its relationship with the external ISM magnetic field lines for the two models with $\alpha=0.95$, being $B_0=5\mu$G (top) and $B_0=0.5$mG (middle). The bottom images represent the projections obtained for model 
  with with $\alpha=0.6$ (elliptical) and $B_0=0.5$mG. The visualizations were obtained for two LOS, the first along z-axis and the second with an angle of 30 degrees.}
  \label{vizualization}
\end{figure*}

The 3-dimensional structure of the nebula is exactly as shown in the two dimensional maps. In order 
to provide the reader a more realistic visualization of the nebula and the environment field lines 
we provide in Figure \ref{vizualization} the projections of the density from models BIP9 and BIP10. 
The figures are simply a volumetric projection of the density distribution along a given line of sight. 
Here we illustrate the projections for a LOS parallel to the z-axis, i.e. perpendicular to both the 
nebula axis of symmetry and the external magnetic field, as well as for a LOS tilted by $30^\circ$. 
The gas distribution is overplot by the magnetic field lines, represented as red tubes. Model BIP9, with 
$B_0=5\mu$G, shows magnetic field lines that are distorted by the expanding shell. The field lines lay 
over the nebula once compressed by the flow, and are stretched into different directions due to the 
gas motions. Naturally, the nebular axis of symmetry is not aligned to the external magnetic field. 
For BIP10, on the other hand, the strong external field of $B_0=500\mu$G is responsible for the 
break of the nebular expansion towards the direction perpendicular to ${\bf B}_0$. The shell therefore 
expands further along the field lines. The final morphology presents itself aligned to the external 
field.

Notice that even in the strongly magnetized case the ambient field lines are perturbed at the shell 
surface. Naturally, this perturbation will propagate outwards, as explained before, as 
magnetosonic and Aflv\'en wave modes. The excitement of such modes is the result of the energy transfer 
from the expanding shell to the ambient field. An important issue on this process is that 
polarization vectors may not trace the original orientation of the external field but 
would rather represent the local orientation of the 
perturbed components as well. Considering that magnetic fields would be responsible for most of the 
dust alignment in evolved PNe we calculate the synthetic polarization distributions 
for both models as follows.

\subsection{Dust polarization}
\label{pol}

The effects of external magnetic fields in shaping the PNe at the
center of the Galaxy could possibly be probed by means of polarization
maps.  Polarization by dust is one of the mechanisms known and it
  has been extensively applied on the mapping of magnetic fields in
  many astrophysical environments. Polarization by dust may occur from
  dust intrinsic emission as well as dust absorption of background radiation.
  The typical small column densities of the dust component in PNe,
  compared to that of the ISM, makes the detection of local
  polarization from PNe and proto-PNe challenging.  Even though, these
  have been reported in some PNe and proto-PNe
  \citep[e.g.][]{scarrot95,su03,jurgenson03,ueta07}. \citet{sabin07}
  also reported polarization measurements from infrared emission of
  dust grains.  Different processes may lead to grain alignment, such
as a strong radiation source or magnetic fiels.  It is interesting
then to determine how polarization maps, if observed, would trace the
relationship between the morphology of the expanding nebula and the
external interstellar fields. Here we consider the magnetic alignment
process only, and neglect the radiation pressure from the central
star.  This assumption is plausible at evolved stages of PNe, though
the radiative alignment may be dominant at the early proto-PN phases.
The physics of grain alignment with ambient magnetic fields is a
complex subject and is not in the scope of this study, therefore we
will perform a simplified calculation of the polarization.

For each cell of the simulated cube the angle of alignment 
($\psi$) is determined by the local magnetic field projected into the plane of 
sky, and the linear polarization Stokes parameters $q$ and $u$ are then given by: 

\begin{eqnarray} \label{eq_qu}
q = \epsilon \rho \cos 2\psi \sin^2 i, \nonumber \\
u = \epsilon \rho \sin 2\psi \sin^2 i,
\end{eqnarray}

\noindent
where $\rho$ is the local density and $i$ is the inclination of the local magnetic field with 
respect to the line of sight. The ``observed'' values of $Q$ and $U$ are obtained by integrating 
$q$ and $u$ along the LOS, respectively. The polarization degree is obtained as 
$p = \sqrt{Q^2+U^2}/I$ and the polarization angle by $\phi = \frac{1}{2}\arctan(U/Q)$. 

\begin{figure}
{\centering
 
{Model: BIP9} 
 
 \includegraphics[width=8.3cm]{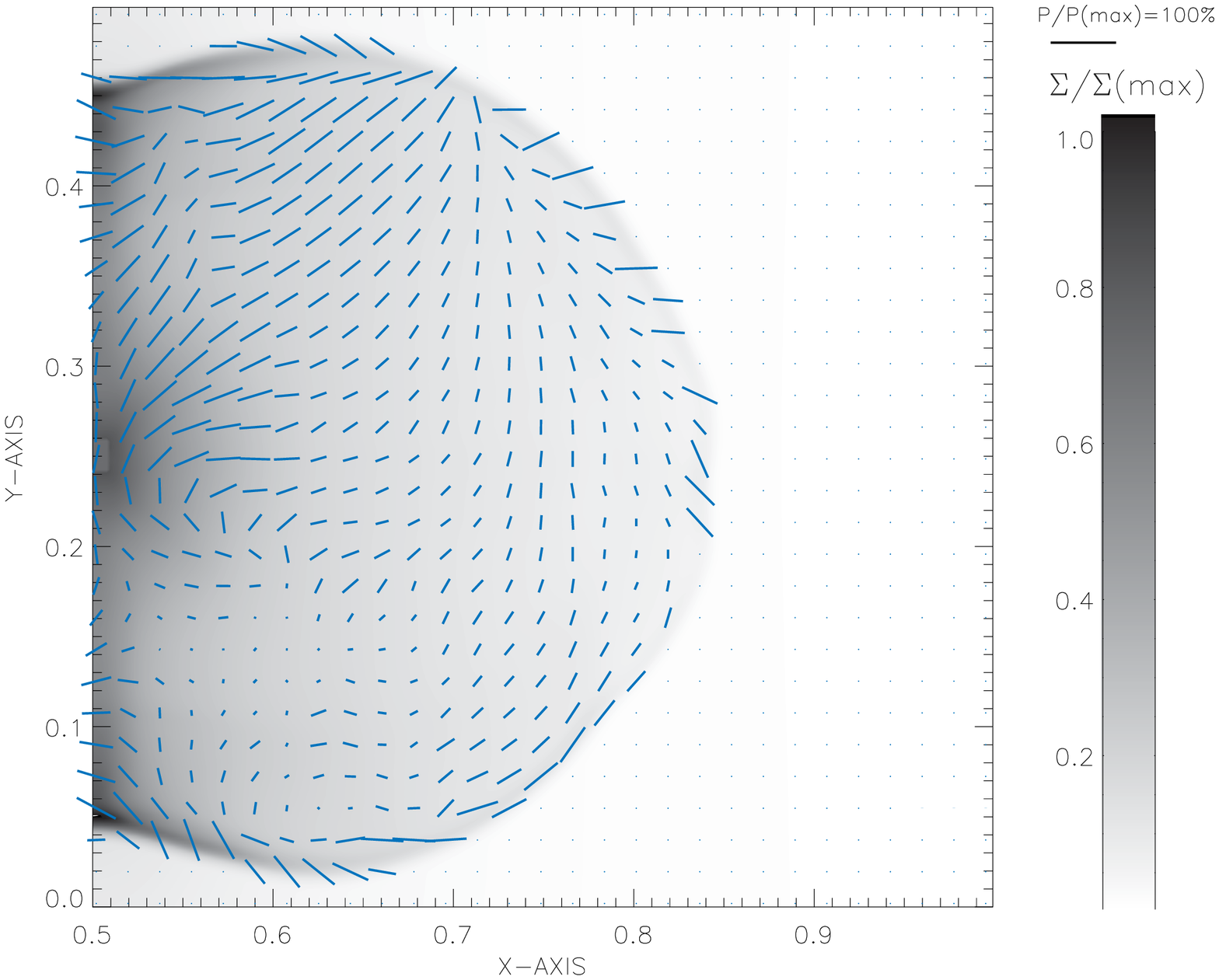} 

{Model: BIP10}

 \includegraphics[width=8.3cm]{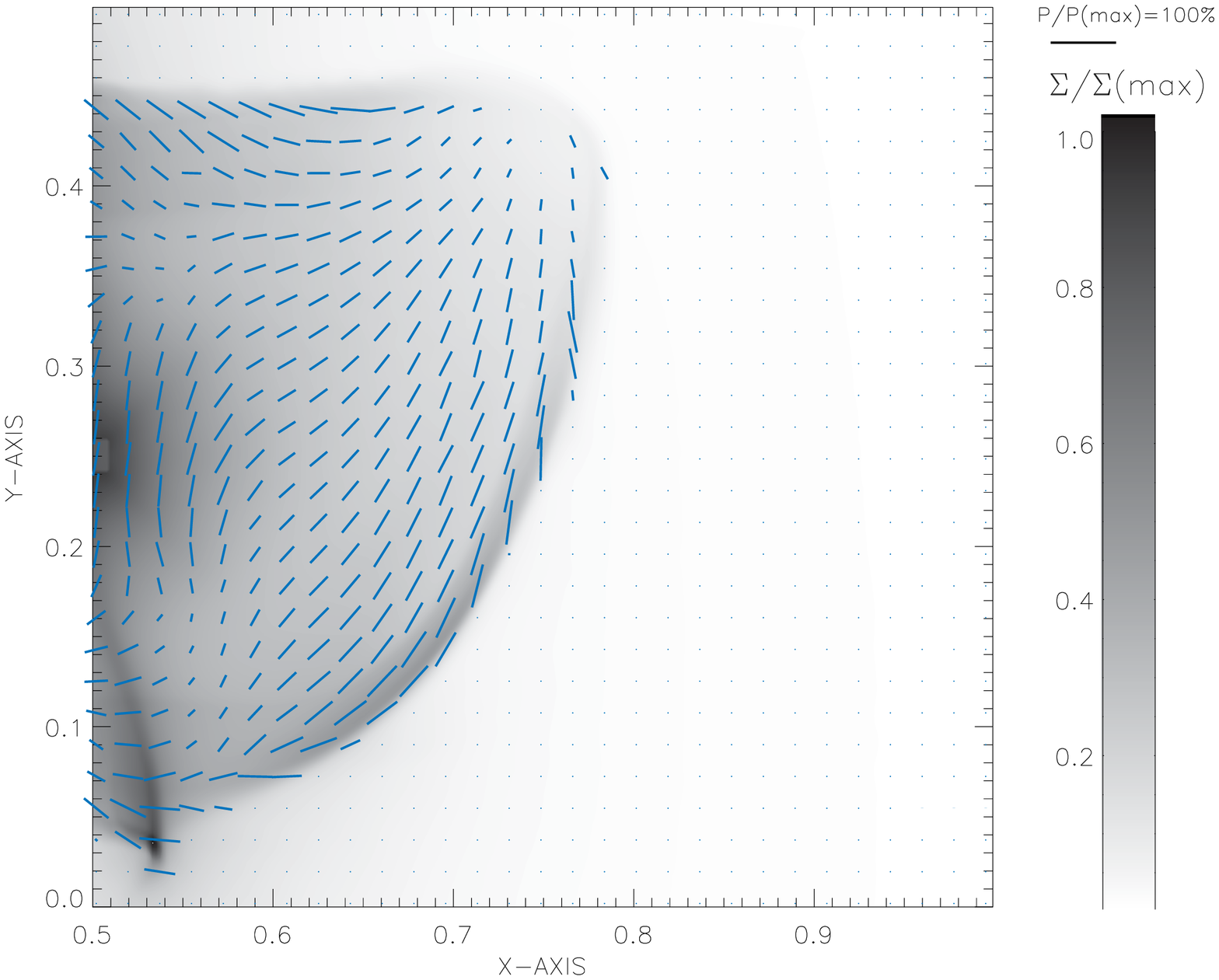}
 
{Model: ELI5} 
 
 \includegraphics[width=8.3cm]{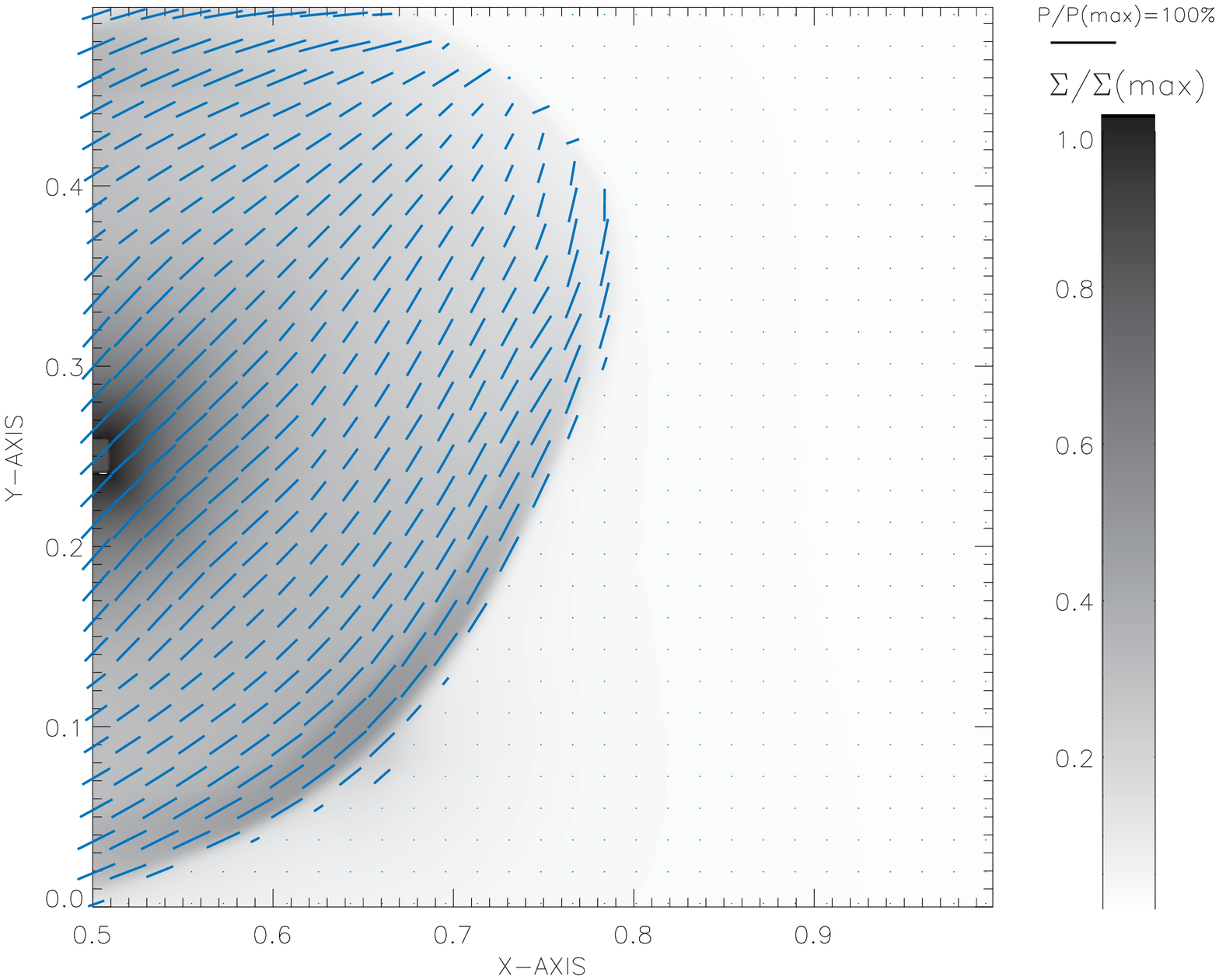}} 
  \caption{Synthetic polarization maps obtained for the models with $\alpha=0.95$ (bipolar), being $B_0=5\mu$G (top) and $B_0=0.5$mG (middle). The bottom image represents the polarization map for model 
  with with $\alpha=0.6$ (elliptical) and $B_0=0.5$mG.}
  \label{polarization}
\end{figure}

In Figure \ref{polarization} we present the synthetic polarization
maps of the two models shown in Figure \ref{vizualization}. The maps
represent the emission measure overplotted by the polarization vectors
expected to be observed in such systems\footnote{Notice that here we
  consider the polarized dust infrared emission and therefore the
  polarization vectors have been rotated by 90 degrees. If one
  considers the dust absorption of background radiation the
  polarization will present a change in relative amplitude and the
  rotatation is not needed.}. We assumed a homogeneous dust alignment
efficiency along the whole nebula.

As mentioned before, based on the 3D visualization of the nebulae, the
polarization vectors for $B_0=500\mu$G (top) are not uniform - as
could be previously thought -, but are also changed by the dynamics of
the nebula. This because the magnetic field probed by the dust
polarization is actually the field within the nebular material,
i.e. which has already interacted with the ejecta. At the edges of the
nebula, the polarization vectors follow the density contours. For
$B_0=5\mu$G (bottom) we obtain a more complex distribtuion of the
plarization vectors.  It is also noticeable that the polarization
intensity is more uniform in the strongly magnetized case compared to
the synthetic map for $B_0=5\mu$G. The main reason for the variation
in polarization degree is the local non-uniformity of the field lines
as these are integrated along the line of sight, which is larger for
the weak magnetic field case.

\section{Discussion and Conclusions}

In this work we studied the dynamical effects of interstellar magnetic
fields on the morphology of PNe. We focus on the late stages of the
nebulae expansion, as we assume that the origin of the intrinsic
morphology of the PNe is unrelated to the large scale interstellar
properties.

In a very simplified analytical approach it is possible to understand
the magnetic pressure at the shell surface as the nebulae expand over
the interstellar medium It is possible to estimate that a PN would
have its morphology modified by the external field once its ram
pressure becomes equivalent to the total piled-up field. This picture
is different from that of models that consider a static ambient
field. For a typical PN, this occurs at $t<10^4$yrs for an ambient
field of $B_0>100\mu$G. Naturally, this is way too strong magnetic
field compared to the estimates of the interstellar medium magnetic
field in the solar neighborhood and in most regions of the Galactic
plane ($\sim 2-5\mu$G) \citep[see][for a review]{beck09}. An exception
to that is the galactic centre/inner bulge. The study of the magnetic
field at the central regions of the Galaxy is challenging, specially
at the disk plane due to sources of contamination and absorption.

 Polarization of radio continuum from the central few hundreds of
  parsecs reveal a dominantly poloidal magnetic field,
  i.e. perpendicular to the disk plane. Magnetic field intensities of
  $100 - 1000\mu$G \citep[see][]{yusef97,chuss03,noutsos12} had been
  inferred.  Lower limits of $> 50\mu$G have also been determined from
  $\gamma$-rays \citep{crocker10} for those regions.  The poloidal
  component is possibly associated to a galactic wind, possibly driven
  by the strong stellar feedback at the central regions of the Milky
  Way. These winds would drag the field lines to the observed
  configuration. The situation is reversed at the disk plane. The ISM
  rotating around the Galactic centre would on the other hand keep a
  dominantly toroidal component, possibly of the same order of
  magnitude. This idea has been corroborated by recent studies that
  reveal a toroidal large scale field with intensities of mG
  \citep{ferr09,nishiyama10}. Therefore, the current understanding is
  that of a two component magnetic field, being one toroidal (mostly
  at the disk plane) and another poloidal, perpendicular to the disk,
  mostly permeating the large height parts of the bulge and the inner
  halo.

The detection of such strong fields reveals the importance of 
  taking into account the interstellar magnetism on (re)shaping
galactic PNe. Despite its low relevance for most of the galactic disk,
the magnetic fields would distort (or bend) nebulae located at the
central regions. This  can explain the divergent conclusions of
\citet{corradi98} with those of \citet{melnick75,phillips97,weid08}
regarding the alignment of PNe with respect to the galactic
disk. Also, it is in agreement with the recent work of \citet{rees13}
that showed  statistically significant alignment of bipolar PNe
at the central region of the Galaxy.  This also  leads to the
conclusion that the magnetic fields are possibly the dominant agent on
the PNe alignment process, and the eventual alignment of the PN with
the disk plane is just the consequence of a magnetic field lying
parallel to the disk plane.  We could also suggest that the
correlation between the alignment of PNe with respect to the disk
plane to be a function of the galactic latitude, even at the galactic
centre. This because the dominant component transits from toroidal to
poloidal with increasing height with respect to the disk plane. 
  However it is difficult to predict at what heights the PNe would
  change from toroidal to poloidally aligned, mostly because of the
  large uncertainties on the observational estimates of ${\bf B}$.

Our analytical estimates were confirmed by a series of numerical
simulations. We perfomed a number of 2.5 and 3-dimensional MHD
simulations in order to verify the validity of the simplified
analytical model. We assume the anisotropic preset density
distribution as the original shaping mechanism of the nebula. Previous
numerical simulations have already tested similar systems \citep[see
e.g.][]{heiligman80,stone92,matt06}, but focused on the original
shaping of the nebula, or assumed parallel symmetries with respect to
the ambient field. Here the models were run with tilted external
magnetic fields, with respect to the original axis of symmetry of the
PN, aiming to study the distortion of evolved PNe by external fields.

We found that typical ISM fields of few $\mu$G are unable to change the dynamics 
of the ejecta. Also, even strong fields are unable to modify the shape of the 
nebulae (e.g. to transform a round nebula into 
a bipolar one). However, we show that strong magnetic fields are able to tilt 
the axis of symmetry of originally aspherical PNe. The alignment of the nebula 
axis of symmetry with respect to the large scale external magnetic field. This 
process is more pronounced in bipolar neabulae, compared to the other morphologies.
This finding is in particular agreement with the findings of \citet{rees13}, 
which showed a preferential alignment for bipolar nebulae as well.  

These results are in agreement with \citet{grinin68} and, since {\bf B} is toroidal at low galactic 
latitudes, i.e. the field lies on the galactic plane, they are also in 
agreement with \citet{melnick75,phillips97,weid08,rees13}. At larger galactic radii and at high 
latitudes {\bf B} is either weak or unrelated to the galactic plane, therefore we consider 
that our results also agree with \citet{corradi98}.
We found good correspondance with the analytical estimates: i) for the minimum magnetic 
field intensity required to dynamically affect the expansion of a typical nebula, as 
$B_0 > 100\mu$G, and ii) found the geometry of the magnetic fields at the PNe to be non-uniform, 
despite of the previous models that assumed static field lines.

One could imagine that polarization maps of planetary nebulae would be
useful in probing the orientation and intensity of the interstellar
field, due to its imprint on the polarization vectors of the nebula
lobes.  To investigate this we calculated synthetic polarization
maps from the 3D MHD models. The maps showed no preferential
alignment, except for a slightly larger polarization degree along the
direction parallel to the external field.  Also a small uniformity of
polarization vectors is seen in that direction. The complexity of the
polarization maps, even in the strongly magnetized case, is related to
the fact that the magnetic energy is concentrated right outside the
dense nebula. The motions in the shell are actually super-Alfv\'enic
and the field lines will be distorted in any case. Since most of the
polarization is due to the denser regions, the intense magnetic field
is not probed.

 Our results also show that galactic PNe could be aligned to the
galactic plane, if the local field is parallel to the galactic disk
and if its intensity is large ($>100\mu$G). The reason for the
alignment is simply the dynamical effect of the magnetic pressure in
modifying the expansion velocities of the ejecta.
 
 Finally, our model predicts that the alignement correlation 
should be size/age-dependant. If the angular momentum of
  the progenitor star is unrelated to the external field, the initial
  orientation of the PNe should be random. Only after expandind to
  quasi-equipartition with the ISM magnetic pressure the nebula would
  be tilted. From our estimates, this should occur around $\sim
  10^3$yrs after ejection.  If we consider a typical expansion
  velocity of few tens of km s$^{-1}$, we obtain angular lobe lengths
  of $l<0.5$arcsec for objects at the central region of the Galaxy.
  Unfortunately, the data of \citet{rees13} is dominated by objects
  larger than 1 arcsec, and this prediction could not be tested. It
  would be interesting for future observational surveys with high
  spatial resolution to explore this predicition.

\section*{Acknowledgments}

DFG thanks the European Research Council (ADG-2011 ECOGAL), and
Brazilian agencies CNPq (no. 300382/2008-1), CAPES (3400-13-1)
and FAPESP (no.2011/12909-8) for financial support. 
HM thanks CNPq grant 573648/2008-5 and FAPEMIG
grants APQ-02030-10 and CEX-PPM-00235-12.

\end{document}